\documentclass[aps,10pt]{revtex4}
\usepackage{graphicx,subfigure}
\usepackage{hyperref}
\usepackage{epsfig}
\usepackage{xcolor}
\usepackage{amsmath}
\usepackage{amssymb}
\usepackage{amsthm}
\usepackage{bbold}
\begin{document}
	\title{Statistical Mechanics of DNA Mutation using SUSY Quantum Mechanics}
	\author{K. Haritha and K. V. S. Shiv Chaitanya }
	\email[]{ chaitanya@hyderabad.bits-pilani.ac.in}
	\affiliation{Department of Physics, BITS Pilani, Hyderabad Campus,  \\Jawahar Nagar, Kapra Mandal, Medchal
		Dist, Hyderabad,\\ Telangana India 500 078.}
	\begin{abstract}
	In this paper, we investigate DNA denaturation through Statistical Mechanics and show that exceptional polynomials lead to DNA mutation. We consider a DNA model with two chains connected by Morse potential representing the H bonds, and then we calculate the partition function for this model.  The partition function is converted into a Schr\"odinger like an equation.  We exploit the techniques of SUSY quantum mechanics to model DNA mutation. We also compute  the thermal denaturation of DNA for each mutated state.
	\end{abstract}
	
	\maketitle
\section{Introduction}	
	DNA (Deoxyribonucleic acid) is a complex molecule composed of two nucleotide chains that form a spiral called a double helix.  In a nucleotide base is attached to a sugar and a phosphate molecule, the base is made up of chemicals adenine (A), guanine (G), cytosine (C), and thymine (T). The sequence of these bases encodes instructions that determine the information available for building and maintaining an organism. DNA is a common molecule that is present in all living organisms. Every cell contains a full set of DNA, which directs the cell about the kind of proteins to be prepared; this, in turn, is responsible for the functioning, growth, and reproduction of all known organisms and many viruses.  An important property of DNA is that it can replicate or make copies of itself. Replication of DNA is critical as the new cells need to have an exact copy of the DNA present in the old cell. In making a copy of itself, a mutation may occur, a permanent change in the nucleotide sequence of DNA. Damaged DNA can be mutated either by substitution, deletion, or insertion of base pairs. The mutation that occurs due to substitution generally happens when the sequence of DNA is altered. The mutation is caused due to deletion that occurs when one or more base pairs are lost. Insertion mutation is the reverse of deletion; that is, it occurs when one or more base pairs are inserted in a DNA sequence\cite{grif}.\\

The significant field of study concerning DNA among the various disciplines of science is thermal denaturation. It refers to the melting of double-stranded DNA into two single strands. The unwinding occurs when hydrogen bonds between the bases in a duplex break due to elevated temperatures. The study of thermal denaturation becomes crucial since it is considered the preliminary step for DNA transcription. Therefore, the DNA helix stability is due to the packing of the bases on top of one another. Therefore, to denature DNA, one must overcome these packing energies providing bonding between adjacent base pairs. Several models have been proposed in the literature to understand DNA denaturation; the original one is using soliton-like solutions. The other popular model is by studying the DNA denaturation from a Statistical mechanics point of view. DNA denaturation and nonlinear excitation of bases are presented as a simple model by Peyrard and Bishop\cite{pet}. This model analyzed the variation of interstrand separation as a function of temperature. In work \cite{Gae} the degrees of freedom of DNA strands and thermodynamics of nonlinear dynamics of DNA was explored. The crucial parameters a, D required to derive the DNA denaturation temperature by statistical methods are determined experimentally in work done by \cite{zdr}
In another work by \cite{guti}, the thermodynamic properties of DNA under the influence of hump-Morse potential are determined. The helicoidal model of the DNA is reviewed by \cite{slob}. In work done by El Kinani \textit{et al},  the q-deformed Morse potential is regarded to describe the DNA denaturation and its statistical properties \cite{kin}.

This paper gives a mathematical model for Statistical Mechanics of DNA mutation using SUSY quantum mechanics. We define three types of mutations, \textit{viz} substitution, deletion, and insertion of base pairs, using SUSY quantum mechanics. The SUSY quantum mechanics can be classified into two types. The first, SUSY Quantum mechanics with shape invariant potentials. The second, SUSY Quantum mechanics without shape invariant potentials. In literature, the second one is defined in terms of codimension exceptional family \cite{gom}.   
SUSY Quantum mechanics with shape invariance arises when two Hamiltonians with potentials, known as partners,  involve ordinary Laguerre/Jacobi or both the solutions involve exceptional Laguerre/Jacobi polynomials.
SUSY Quantum mechanics without shape invariance occurs when the two Hamiltonian with potentials one partner having the solution as ordinary Laguerre/Jacobi and the other having exceptional Laguerre/Jacobi polynomial \cite{Quesne}. For a review of SUSY and more details, refer the reader to \cite{khareajp} and \cite{kharebook}.
In supersymmetry, the superpotential $\mathcal{W}(x)$ is defined in terms of the intertwining operators $ \hat{A}$ and $\hat{A}^{\dagger}$  as
	\begin{equation}
	\hat{A} = \frac{d}{dx} + \mathcal{W}(x), \qquad \hat{A}^{\dagger} = - \frac{d}{dx} + \mathcal{W}(x).
	\label{eq:A}
	\end{equation}
	This allows one to define a pair of factorized Hamiltonians $H^{\pm}$ as
	\begin{eqnarray}
	H^{+} &=& 	\hat{A}^{\dagger} \hat{A} 	= - \frac{d^2}{dx^2} + V^{+}(x) - E, \label{vp}\\
	H^{-} &=& 	\hat{A}  {\hat A}^{\dagger} 	= - \frac{d^2}{dx^2} + V^{-}(x) - E, \label{vm}
	\end{eqnarray}
	where $E$ is the factorization energy. 
	
	The partner potentials $V^{\pm}(x)$ are related to $\mathcal{W}(x)$ by 
	\begin{equation}
	V^{\pm}(x) = \mathcal{W}^2(x) \mp \mathcal{W}'(x) + E, \label{gh}
	\end{equation}
	where  prime denotes differentiation with respect to $x$. The eqs.(\ref{vp}) and (\ref{vm}), imply
	\begin{equation}
	H^{+} \hat{A}^{\dagger} = \hat{A}^{\dagger} H^{-}, \qquad \hat{A} H^{+} = H^{-} \hat{A}.\label{df}
	\end{equation}
	From the above, one can see that the operators $\hat{A}$ and $\hat{A}^{\dagger}$ act as intertwining operators. These allow one to go from one wave function $\vert\psi^{+}_{\nu} \rangle$ to the other $\vert\psi^{-}_{\nu} \rangle$ and vice versa.
	
	The discovery of new exceptional polynomials in the last decade by G\'omez Ullate \textit{et al} has given rise to a rich structure in SUSY quantum mechanics. The classification of exceptional orthogonal polynomials is given in terms of codimension exceptional family \cite{gom}.  Codimension of the exceptional family is defined as the total number of the missing degree in the polynomial sequence. These missing degrees in the polynomial put constraints on the Sturm-Liouville equation and, in turn, on the Bochner theorem. They have also shown that every exceptional orthogonal polynomial system is related to the respective classical orthogonal polynomials by a sequence of Darboux transformations. Further, every exceptional orthogonal polynomial system is related to classical orthogonal polynomials by a sequence of Darboux transformations.  Gomez Ullate \textit{et al} showed that the exceptional orthogonal polynomial systems emerge as eigenfunctions of Sturm-Liouville problems \cite{gom}, \cite{gom1}. The reference \cite{kvs} shows that the quantum mechanical problem, which admits classical Laguerre/ Jacobi polynomials as solutions for the Schr\"odinger equations,  will also admit exceptional Laguerre/ Jacobi polynomials as solutions having the same eigenvalues. Still, with the ground state missing after a modification of the potential. These new polynomials can be classified based on codimension.  Quesne constructed the Darboux transformation for the family of polynomials of codimension-2 \cite{Quesne, Quesne1}. Later Oodake and Sasake constructed exceptional polynomials for polynomials of arbitrary codimension \cite{odake1, odake}.
	In an earlier work done by \cite{kvs2}, the application of exceptional potentials to the relativistic hydrogen atom and the Dirac oscillator has led to some exciting results.  In our case, both the Hamiltonians' wave functions are known, and one can construct the operator $\hat{A}$ from the following relation
	\begin{equation}
	\hat{A}H^{+}\vert\psi^{+}_{\nu} \rangle=E_\nu \hat{A}\vert\psi^{+}_{\nu} \rangle
	=E_{\nu+1}\vert\psi^{-}_{\nu+1} \rangle=H^{-}\vert\psi^{-}_{\nu+1} \rangle\label{oper}
	\end{equation}
	using the operator $\hat{\cal{O}}$   \cite{gom, gom1, Quesne1}, which connects the ordinary Laguerre polynomials to the exceptional Laguerre
	polynomials 
	\begin{equation}
	\hat{\cal{O}} L_{\nu}^{k-1}(x)={\cal L}^{k}_{\nu+1}(x),    \label{34}
	\end{equation}
	where $\hat{\mathcal{O}}=(x+k)(\frac{d}{dx}-1)-1$. The superpotential, ${\cal W}(x)$, can be  obtained 
by replacing $\frac{d}{dx}$ in  $\hat{A}$ in terms of $\hat {\cal O}$. The superpotential ${\cal W}(x)$ is determined. 
Therefore, this should be clear to the readers that SUSY quantum mechanics has two Hamiltonians known as partners, mathematically related by an intertwining operator. In quantum mechanics, these two Hamiltonians correspond to two different states of the system; for instance, one is the ground state and another, the first excited state. But, when we study the Statistical Mechanics of DNA, the partition function of the DNA model with Morse potential is converted into a Schr\"odinger like equation using integral transfer method; here, no quantum mechanics is involved. Hence, there are no different quantum states. So the two partner Hamiltonians should correspond to the different sequencing of the bases chemicals bonded by hydrogen atoms. Therefore, we define substitution DNA mutation in SUSY quantum mechanics when the two partner potentials are related by shape invariance. The deletion mutation occurs when two partner potentials are not related by shape invariance is missing degrees of polynomials.
 
\section{The simple model of DNA Mutation} 
We consider a DNA model with transverse displacement in an nth nucleotide where one strand is indicated as $u_n$, and another chain is indicated as $v_n$. For simplicity, we assume a harmonic potential connects the two neighboring nucleotides of the same strand. For those belonging to different strands, Morse potential was used \cite{pet} to represent the transverse interaction of the bases in a pair. 
The model includes two degrees of freedom corresponding to the two bases of the base pair. It describes the
hydrogen bonds,
The in-phase and out of phase coordinates are calculated
 \begin{equation} 
 x_n=(u_n+v_n)/\sqrt {2}, y_n=(u_n-v_n)/\sqrt{2}
 \end{equation}
  A Hamiltonian is derived assuming a harmonic coupling between the neighboring bases. 
\begin{equation}
H= \Sigma \frac{1}{2}m(\dot u_n^2-\dot v_n^2)+\frac{1}{2}k[(u_n-u_{n-1})^2+(v_n-v_{n-1})^2]+V(u_n-v_n)
\end{equation}
m being the reduced mass of the bases and Morse potential, the potential between base pairs which is given by 
\begin{equation}
V(u_n-v_n) = D[exp[-a(u_n-v_n)]-1]^2
\end{equation}
This paper aims to model DNA Mutation and compute the thermal denaturation of DNA for each mutated state; hence, we use statistical mechanics for this purpose. The classical partition function in the canonical ensemble is constructed in terms of the Hamiltonian having four terms.
\begin{equation}
Z= \int_{-\infty}^{+\infty}\prod_{n=1}^{N} e^{-\beta H(x,y,P_x,P_y)}dx_ndy_ndp_ndq_n
\end{equation}
\begin{equation}
Z=Z_xZ_pZ_yZ_q
\end{equation}
The partition function can be decoupled, and the momentum terms are Gaussian integrals and can be integrated to yield
\begin{equation}
Z_p=Z_q=(2\pi mk_BT)^{N/2} \label {zp}
\end{equation}

The X-coordinate partition function can also be readily integrated which yields\\
\begin{equation}
Z_x=(\frac{2\pi k_BT}{K})^{N/2} \label{zx}
\end{equation}
The Y-coordinate partition function involves the potential term. Nearest neighbor interactions and one-dimensional system are considered, 
\begin{equation}
Z_y=\int_{-\infty}^{+\infty} \prod_{n=1}^{N}  e^{-\beta f(y_n,y_{n-1})} dy_n \label{zy}
\end{equation}
	Where 
\begin{equation}
f_{(n,n-1)}=\frac{k}{2}(y_n-y_{n-1})^2+V_q(y_n)
\end{equation}
where f represents the potential energy component of $H_y$.
This integral is solved exactly in the limit of a large thermodynamic system(N$ \to$$\infty$) using the transfer integral (TI) technique let us operate the transfer integral operator $y_n\rightarrow y_{(n-1)}$ and whose eigen functions are defined by
\begin{equation}
Z_y=\int dy_{n-1}e^{-\beta f(y_n,y_{n-1}) }\phi_i(y_{n-1})=e^{-\beta\epsilon i}\phi_i(y_n)\label{part}
\end{equation}
An equation identical to Schr\"odinger equation is obtained as was solved by \cite{krum},\cite{Curr}\\ 
 
 \begin{equation}
 -\frac{1}{2\beta^2k} \frac{\partial ^2\phi_i(y)}{\partial y^2 }+V(y)\phi_i(y)=(\epsilon_i-s_o-D)\phi_i(y) \label{sch}
 \end{equation}
 the Morse potential given by
 \begin{equation}
 V(y)=V_0(exp(-2\beta y)-2\exp(-\beta y)),
 \end{equation}
here we take $ \beta =2a$ and $V_0= D$. To solve the Schr\"odinger equation (\ref{sch}) we use SUSY quantum mechanics method given in  \cite{khareajp, kharebook}  and the superpotential for the Morse oscillator is given by
 \begin{equation}
{\cal W}(x)= a-be^{(-\alpha x)}
 \end{equation}
 then the potential is given by
 \begin{equation}
 V_o(a,x)= W^2(x,a)-W'(x,a)
  = a^2-2b(a+\frac 1 2 \alpha)e^{-\alpha x}+b^2e^{-2\alpha x}\label{p1}.
 \end{equation}
 We obtain the  partner potential by scaling $a$ to $a-\alpha$ 
 \begin{equation}
 V_1(x,a)= W^2(x,a)+W'(x,a)
  = a^2-2b(a-\frac 1 2\alpha)e^{-\alpha x}+b^2e^{-2\alpha x}\label{p2}
 \end{equation}
 Thus the eigenfunctions are computed as follows
 \begin{equation}
 \psi_0^{(0)}(x,a)=(\frac{2b}{\alpha}e^{-\alpha x})^\frac a {\alpha}e^{-\frac1 2\frac {2b}{\alpha}e^{-\alpha x}}=y(x)^\frac a{\alpha}e^{-\frac 1 2 y(x)}
 \end{equation}
 and the first exited is computed as
 \begin{eqnarray}
 \psi_1^{(0)}(x,a)&=&(E_1^(0)-E_0^0)^-{\frac 1 2}A*(x,a)\psi_0^(0)(x,f(a))\nonumber\\
 &=&(\frac{2a}{\alpha}-1)^{-\frac1 2}y(x)^{\frac a {\alpha}}e^{-\frac 1 2 y(x)}\left(\frac{2a}{\alpha}-1- y(x)) \right)
 \end{eqnarray}
 In general any state is computed as
 \begin{eqnarray}
 \psi_m^{(0)}(x,a)&=&\prod_{i=1}^{m}\left((E_m^0-E_{i-1}^0)^{\frac 1 2}\left(-\frac d {dx}+W(x,f^{(i-1)}a)\right)\right)\psi_0^{0}(x,a-\alpha m)\\
&=&  y(x)^{\frac a {\alpha}-m}e^{-\frac1 2 y(x)}L_m^{2(\frac a {\alpha}-m)}(y(x))
 \end{eqnarray}
 where 
 \begin{equation}
 E_m^{(0)}= \sum_{k=1}^{m}(a-(k-1)\alpha)^2-(a-k\alpha)^2=a^2-(a-m\alpha)^2 
 \end{equation}
 with $y(x)=\frac{2b}{\alpha}e^{-\alpha x}$ and $L_n^{\alpha}(x) $ the Laguerre polynomials \cite{khareajp, kharebook}. 
 In our case $a^2=b^2=D$ and $\alpha=\beta$, hence the partition function (\ref{part}) becomes
 \begin{equation}
Z_y=\sum_{i=1}^{m} \int dy_{n-1}e^{-\beta \left((E_m^0-E_{i-1}^0)^{\frac 1 2}\left(-\frac d {dy}+W(y,f^{(i-1)}a)\right)\right)} \phi_i(y_{n-1})=e^{-\beta\epsilon i}\phi_i(y_n)\label{part1}
 \end{equation}
 The eigenvalues are given by
 \begin{equation}\epsilon_o = \frac{1}{2\beta}ln \frac{\beta k} {2\pi}+\frac{a}{\beta}\sqrt{\frac{2D}{k}}-\frac{a^2}{2\beta^2 k }\label{e0}
 \end{equation}
 and
 \begin{equation}
 s_o=\frac{1}{2\beta}ln \frac{\beta k} {2\pi}. \label{s0}
 \end{equation}
 The normalized eigenfunction for the ground state is calculated to be
 \begin{equation}
 \phi_0(y)= (\sqrt2a)^{1/2}\frac{(2d)^{(d-1/2)}}{[\Gamma(2d-1)]^{1/2}} exp(-de^{\sqrt2a y}) exp(-d-1/2)\sqrt2 ay.\label{gs}
 \end{equation}
 We compute the free energy for the ground state and is given by
 \begin{equation}
 f=  \epsilon-K_bT [N ln (2\pi mK_bT) +\frac N 2 ln(\frac {2mK_bT} K ) ]
 \end{equation}
 where free energy is defined by
 \begin{equation}
 f= -K_bT ln( Z_pZ_q Z_xZ_y).
 \end{equation}
  The free energy of the first excited state of the ordinary potential is given by
  \begin{eqnarray}
 f&=& \epsilon  ln\left[\frac{(\sqrt2a)^{\frac 3 2} (2d)^{d-\frac{3}{2}}}{[(2d-2)(2d-3)\Gamma(2d-3)]^{\frac 1 2}}
 exp(-de^{-\sqrt 2 ay})exp(-(d-\frac{1}{2})\sqrt 2 ay)\left(1-\sqrt2dexp(-2ay)\right)\right]\\&& \nonumber
 -NK_bT ln \frac{(2\pi mK_bT)^\frac 3 2}{\sqrt \pi K}
 \end{eqnarray}	
 where the first excited eigenfunction of the Morse potential is given by
 \begin{equation}
 \phi_0^1= (\sqrt2a)^{3/2} \frac{(2d)^{(d-3/2)}}{[(2d-2)(2d-3)\Gamma{(2d-3)}]^{1/2}} exp(-de^{\sqrt2 ay})exp[-(d-1/2)\sqrt2ay](1-\sqrt2dexp(-2ay))\label{fs}
 \end{equation}
We compute the denaturation temperature as average stretching of base pairs with respect to temperature for different values of coupling constants by plotting the  $< y >$ given by
\begin{equation}
<y>=\int y \phi_i(y)dy
\end{equation}
with temperature by fixing the values of
$\beta=\frac{1}{K_BT}$, $a=1.8 \dot{A}^{-1}$ and force constant to be
\begin{equation}
K=2X10^3eV/\dot A^2, 3X10^{-3}eV/\dot A^2, 4X10^{-3} eV/\dot A^2.
\end{equation}
 
In the figure FIG1,  we have plotted $< y >$ the variation of average stretching of base pairs with respect to temperature for the ground state wave function (\ref{gs}) with different values of coupling constant. 
 \begin{figure}[hbt!]
 	\centering
 	\includegraphics[width=8cm]{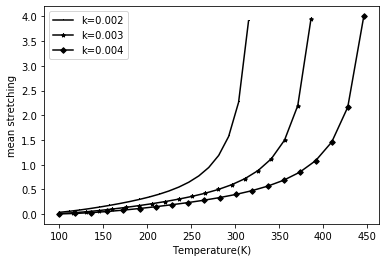}
 	\caption{Variation of $<y>$ as a function of Temperature for three values of coupling constant K!}
 \end{figure}
%\begin{figure}[hbt!]
%	\centering
%	\includegraphics[width=8cm]{ysq-normal.PNG}
%	\caption{Variation of $<y^2>$ as a function of Temperature for three values of coupling constant K!}
%\end{figure}
In the figure FIG2,  we have plotted $< y >$  the variation of average stretching of base pairs with respect to temperature for the first exited state wave function (\ref{fs}) with different values of coupling constant.
\begin{figure}[hbt!]
	\centering
	\includegraphics[width=8cm]{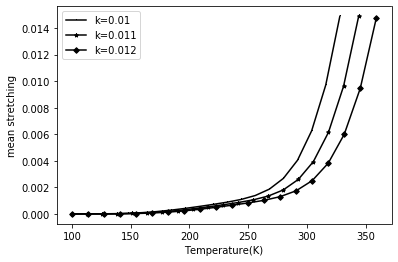}
	\caption{Variation of $<y>$ as a function of Temperature for three values of coupling constant K!}
\end{figure}
%\begin{figure}[hbt!]
%	\centering
%	\includegraphics[width=8cm]{ysq-fes.PNG}
%	\caption{Variation of $<y>^2$ as a function of Temperature for three values of coupling constant K!}
%\end{figure}

For the first excited state, the K values that are used are 
$K=10X10^{-3}, 11X10^{-3}$  and $ 12X10^{-3}$
 and $a=0.8 A^{-1}$
  It is clear from the equation (\ref{part1}) when $i=1$ one gets the ground state and $i=2$ corresponds to first excited state and so on. The mutation that occurs due to substitution generally happens when the sequence of DNA is altered. From the partition function (\ref{part1}), it is clear that if we change the value of $i$ from $1$ to $2$, one goes from the ground state to the first excited state. When the DNA makes a change from the ground state to the first excited state, the wave function changes, so does the potential.  But from equations (\ref{p1}) and (\ref{p2}) it is clear that the shape of potential is unchanged and differs by a constant. Hence this does not change the Schr\"odinger equation (\ref{sch}). Therefore, we conclude a mutation occurs when DNA undergoes a transition from the ground state to the first excited state. The change in the eigenfunction of the Schr\"odinger equation (\ref{sch}) corresponds to altering the sequence in DNA, which in turn reflects in the chemical bonds. Thus we conclude this corresponds to a mutation, namely substitution.

 \section{Mutation by Deletion}
 The deletion mutation occurs when one or more base pairs are lost. To model deletion mutation using SUSY quantum mechanics where the solution to the Schr\"odinger equation (\ref{sch}) are exceptional polynomials. One partner potential corresponds to the Laguerre polynomials in the SUSY quantum mechanics solution, and another partner potential corresponds to the exceptional Laguerre polynomials. The Schr\"odinger equation (\ref{sch}) goes from Laguerre polynomials to the exceptional Laguerre polynomials. The ground state is dropped; this corresponds to a loss of nucleotide or base pairs; hence this mutation corresponds to Deletion. Consider the Schr\"odinger equation (\ref{sch}), by making the following change of variable
 \begin{equation}
 z= \sqrt2  d exp(-2ay)
 \end{equation} 
 and $\rho=\frac{z}{n}$, then
 the  Schr\"odinger equation (\ref{sch}) reduces to
 the associated Laguerre differential equation given by 
 \begin{equation}
 \rho F''(\rho)+[2s+1-\rho] F'(\rho)-(s+\frac{1-\sqrt 2d}{2}) F(\rho) =0. \label{lag}
 \end{equation}
 and the solutions are Laguerre polynomials and are given by 
 \begin{equation}
 \phi_n(\rho)= N_n^\nu e^{-\frac{\rho}{2}}\rho_n^sL_n^{2s}(\rho)
 \end{equation}
 where eigenvalues are given in equation (\ref{e0}) and (\ref{s0}), 
 $\nu=\sqrt{2}d$, and 
 $s= \frac{\nu-1} 2$.
 In order to find the exceptional Laguerre polynomial as the solution to the Schr\"odinger equation (\ref{sch}) we use the following theorem  \cite{kvs}:
  By adding an extra term $V_e(x)$ to the Laguerre/ Jacobi differential equation and demanding the solutions to be  $g(x)=\frac{f(x)}{(x+m)}$ and $g(x)=\frac{f(x)}{(x-b)}$
 for the new differential equations, where $f(x)$ is the Laguerre and Jacobi polynomials respectively. Then $g(x)$ satisfies $X_1$- exceptional differential equation for the Laguerre and Jacobi respectively, $V_e(x,m)$ can be determined uniquely. that is by adding the extra potential $V_e$ to the Laguerre equation \ref{lag}  
 \begin{equation}
 zH_n''(\rho)+(2s+1-\rho)H_n'(\rho)+(n+V_e) H_n(\rho)=0\label{ae}
 \end{equation}
setting $2s=m $ and $n=\lambda-1$ and by demanding the solution to the differential equation to be
 \begin{equation}
H(\rho)=\frac{f_n(\rho)}{\rho+m}
\end{equation}
 where $H(\rho)$ satisfies the $X_1$ exceptional Laguerre differential equation 
 \begin{equation}
 -x f''(\rho) +\left ( \frac{\rho-m}{\rho+m} \right ) [(m+\rho+1)f'(\rho)-f(\rho)]=(n-1)f(\rho), \label{exlau}
 \end{equation} 
 and the eigenfunction is given by 
 \begin{equation}
 \phi_o(z)=\left[\frac{1}{\Gamma(2d-1)}\right]^\frac{1}{2} (\sqrt2a)^{\frac 1 2}(2d)^{d-\frac 1 2}e^{({-(d-\frac 1 2)}{\sqrt 2 ay} )}exp(-de^{-\sqrt 2ay}){\cal L}_n(z)
 \end{equation}
 where $s=\frac{\sqrt2d-1}{2}$
 \begin{equation}
 \phi_0^e(z)=\frac{(\sqrt2a)^{\frac 1 2}}{\sqrt2d-1+\sqrt 2dexp(-2ay)}\frac{(2d)^{{d-\frac1 2}}}{[\Gamma(2d-1)]^{\frac 1 2}}exp(-de^{-\sqrt 2 ay})exp[-(d-\frac{1}{2})\sqrt 2 ay]{\cal L}(y)\label{gs1}
 \end{equation}
which determines $V_e(\rho,m)$ to be 
 \begin{equation}
 V_e(\rho,m)= \frac{2m}{(\rho+m)^2}-\frac{1}{\rho+m}.
 \end{equation}
 Substituting $m=2s$
 we obtain the exceptional partner potential as
 \begin{equation}
 V_e(z,s)= \frac{4s}{(z/n+2s)^2}-\frac{1}{z/n+2s}.
 \end{equation}
 We see that the potential depends on the quantum number $n$ and
 thus the Morse potential is a conditionally or quasi exactly solvable model. Recently in reference \cite{q1} has shown that  Morse potential is a  quasi exactly solvable model. These models were discovered in \cite{cond}.  We believe that in our model for DNA, the conditionally exactly solvable model makes sense as the new state of DNA that is exceptional Laguerre polynomials has the signature of the old state that is Laguerre polynomials. The free energy per cite for the new state of DNA with exceptional potential is given by
 \begin{equation}
f= -K_bT\left [N ln (2\pi mK_bT) +\frac N 2 ln(\frac {2mK_bT} K ) + [ \sum_{n=1}^{N}  {-\beta\epsilon_i}ln(\phi_i(y_n))]\right].
\end{equation}
Considering the ground state for the  exceptional potential that is $ \phi_o^e(z)$ and free energy is given by
\begin{eqnarray}
f&=& \epsilon  ln\left[\frac{(\sqrt2a)^{\frac 1 2}}{2d-1+\sqrt2dexp(-2ay)}\frac{(2d)^{d-\frac{1}{2}}}{[\Gamma(2d-1)]^{\frac 1 2}}
exp(-de^{-\sqrt 2 ay})exp(-(d-\frac{1}{2})\sqrt 2 ay)\left(1-\sqrt2dexp(-2ay)\right)\right]\\&& \nonumber
-NK_bT ln \frac{(2\pi mK_bT)^\frac 3 2}{\sqrt \pi K}
\end{eqnarray}	

We compute the denaturation temperature as average stretching of base pairs with respect to temperature for different values of coupling constants by plotting the  $< y >$ given by
\begin{equation}
<y>=\int y \phi_0^e(y)dy
\end{equation}
with temperature by fixing the values of
$\beta=\frac{1}{K_BT}$, $a=1.8 \dot{A}^{-1}$ and force constant to be
\begin{equation}
K=2X10^3eV/\dot A^2, 3X10^{-3}eV/\dot A^2, 4X10^{-3} eV/\dot A^2.
\end{equation}
In the figure FIG3,  we have plotted $< y >$  the variation of average stretching of base pairs with respect to temperature for the exceptional potential ground state (\ref{gs1}) with different values of coupling constant which are identical with the values of ground state wave function (\ref{gs}). In this case we observe that the exceptional potential denaturation of DNA strands occurs at lower temperatures when compared to the normal potential.
%\begin{figure}[hbt!]
%	\centering
%	\includegraphics[width=8cm]{7.png}
%	\caption{Variation of $<y>$ as a function of Temperature for three values of coupling constant K!}
%\end{figure}
%\begin{figure}[hbt!]
%	\centering
%	\includegraphics[width=8cm]{8.png}
%	\caption{Variation of $<y^2>$ as a function of Temperature for three values of coupling constant K!}
%\end{figure}
In the figure FIG3 and FIG4,  we have plotted $< y >$ and $< y^2 >$ respectively, the variation of average stretching of base pairs with respect to temperature for the exceptional potential ground state (\ref{gs1}) with different values of coupling constant that is $ 4X 10^{-3},5X 10^{-3}, 6X 10^{-3}$ then we note that the denaturation temperature approaches room temperature value.
A graph is drawn between $<y>$ and Temperatures for different values of K
\begin{figure}[hbt!]
	\centering
	\includegraphics[width=8cm]{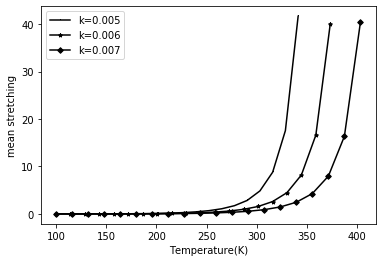}
	\caption{Variation of $<y>$ as a function of Temperature for three values of coupling constant K!}
\end{figure}
Another graph between $<y^2>$ and temperature is drawn.
%\begin{figure}[hbt!]
%	\centering
%	\includegraphics[width=8cm]{ysq-excep.png}
%	\caption{Variation of $<y^2>$ as a function of Temperature for three values of coupling constant K!}
%\end{figure}
From the above discussion we conclude that when the solution for the DNA Hamiltonians are Laguerre polynomials and exceptional Laguerre polynomials respectively, a deletion mutation occurs when DNA undergoes a transition from the  ground state (\ref{gs})  to the ground state (\ref{gs1}). In this case,  the wave functions for both the Hamiltonians are known and one can construct the operator $\hat{A}$ from the following relation 
\begin{equation}
\hat{A}H^{+}\vert\psi^{+}_{\nu} \rangle=E_\nu \hat{A}\vert\psi^{+}_{\nu} \rangle
=E_{\nu+1}\vert\psi^{-}_{\nu+1} \rangle=H^{-}\vert\psi^{-}_{\nu+1} \rangle\label{oper}
\end{equation}
using the operator $\hat{\cal{O}}$   \cite{gom, gom1, Quesne}, which connects the ordinary Laguerre polynomials to the exceptional Laguerre
polynomials 
\begin{equation}
\hat{\cal{O}} L_{\nu}^{k-1}(x)={\cal L}^{k}_{\nu+1}(x),    \label{34}
\end{equation}
where $\hat{\mathcal{O}}=(x+k)(\frac{d}{dx}-1)-1$. We also conclude that the addition mutation is converse of the deletion mutation that is applying $\hat{\cal{O}}^\dagger$ on exceptional Laguerre polynomials. As one can see that the new state is the ordinary Laguerre polynomials where a new state is added hence a addition mutation.

\section{Conclusion}
In this paper, we have investigated DNA denaturation through Statistical Mechanics and shown that exceptional polynomials lead to DNA mutation. The DNA model with two chains connected by Morse potential representing the H bonds, and then we calculate the partition function for this model. By converting the partition function into a Schr\"odinger like equation and exploiting SUSY quantum mechanics techniques, we have given a mathematical model for DNA mutations. Then we computed the free energy and the thermal denaturation of DNA for each mutated state. We have shown a mutation occurs when DNA undergoes a transition from the ground state to the first excited state. The change in the eigenfunction of the Schr\"odinger equation (\ref{sch}) corresponds to altering the sequence in DNA, which in turn reflects in the chemical bonds. Thus we conclude this corresponds to a mutation, namely substitution. We have also shown that the deletion mutation occurs when DNA undergoes a transition from the  ground state (\ref{gs})  to the ground state (\ref{gs1}), and it can be achieved through an operator  $\hat{\cal{O}}$  which connects the ordinary Laguerre polynomials to the exceptional Laguerre polynomials 
\begin{equation}
\hat{\cal{O}} L_{\nu}^{k-1}(x)={\cal L}^{k}_{\nu+1}(x),    \label{38}
\end{equation}
where $\hat{\mathcal{O}}=(x+k)(\frac{d}{dx}-1)-1$. We also have shown that the addition mutation occurs DNA undergoes a transition from the ground state (\ref{gs1})  to the ground state (\ref{gs}). It can be achieved through an operator  $\hat{\cal{O}}^\dagger$ on exceptional Laguerre polynomials.  one can see that the new state is the ordinary Laguerre polynomials where a new state is added hence an addition mutation.
We have shown the dependence of denaturation temperature on the interstrand separation and found that the curve has a non-linear rise far before the rise occurred. The application of exceptional potential involves the dismissal of the ground state of the DNA. Theoretically, there is a mutation of DNA involved in this process, and wherein there is an alteration in the base units of DNA. The mean stretching of hydrogen bonds $<y>$ of the ground state and the first excited state of the eigenfunction have been determined. We have calculated the mean stretching of a new state. We emphasize that though it is not an excited state of the earlier state, it is a new state whose ground state is eliminated due to exceptional potential. We have plotted the dependence of mean stretching of hydrogen bonds corresponding to a new state on temperature and found that the denaturation occurs only for higher coupling constants' values. The role of exceptional polynomials is to delete the ground state of the function. Hence we claim the occurrence of Mutation in the DNA.

\section*{Acknowledgments}
KVSSC acknowledges the Department of Science and Technology, Govt of India ( (D. O. No: MTR/2018/001046)), for financial support.

\end{document}